\documentclass[twocolumn,superscriptaddress,showpacs,pra]{revtex4}
\usepackage{epsfig}
\usepackage{epstopdf}
\usepackage{delarray}
\usepackage{amsmath, amssymb}
\usepackage{bm}
\usepackage{graphicx}% Include figure files
\usepackage{placeins}
\usepackage{color}
\begin{document}
\title{Rogue Quantum Harmonic Oscillations}

\author{Cihan Bay\i nd\i r}
\email{cihanbayindir@gmail.com}
\affiliation{Associate Professor, Engineering Faculty, \.{I}stanbul Technical University, 34467 Maslak, \.{I}stanbul, Turkey. \\
						 Adjunct Professor, Engineering Faculty, Bo\u{g}azi\c{c}i University, 34342 Bebek, \.{I}stanbul, Turkey. \\
						 International Collaboration Board Member, CERN, CH-1211 Geneva 23, Switzerland.}

%\date{\today}
\begin{abstract}
We show the existence and investigate the dynamics and statistics of rogue oscillations (standing waves) generated in the frame of the nonlinear quantum harmonic oscillator (NQHO). With this motivation, in this paper we develop a split-step Fourier scheme for the computational analysis of NQHO. We show that modulation instability excites the generation of rogue oscillations in the frame of the NQHO. We also discuss the effects of various parameters such as the strength of trapping well potential, nonlinearity, dissipation, fundamental wave number and perturbation amplitude on rogue oscillation formation probabilities. 
\end{abstract}
\maketitle

%%%%%%%%%%%%%%%%%%%%%%%%%%%%%%% main %%%%%%%%%%%%%%%%%%%%%%%%%%%%%
\section{\label{sec:level1} Introduction}

Quantum harmonic oscillator (QHO) in quantum mechanics is analogous to the simple harmonic oscillator of the classical vibration theory \cite{Schrodinger, Griffiths, Liboff, Pauli, Messiah}. It is commonly used as a model to study the vibrations of the atomic particles and molecules under effect of classical spring like potential which is a commonly accepted model for the molecular bonding. QHO is one of the exactly solvable models in the field of quantum mechanics having solutions in the form of Hermite polynomials and it can be generalized to N-dimensions \cite{Schrodinger, Griffiths, Liboff, Pauli, Messiah}.

On the other hand, studies on nonlinear quantum mechanical phenomena became popular in the last few decades \cite{Nattermann, Reinisch, Wang_Nl_entang, Bay_Zeno, Bay_TWMS2017, Chia_Vedral}. Majority of the nonlinear quantum mechanics studies rely on dynamic equations, such as the nonlinear Schrödinger equation (NLSE). Starting from the Maxwell's equations for electromagnetism, the NLSE and some of its extensions can be derived for modeling the wavefunctions under the effects of high order nonlinear electric and magnetic fields. Under such fields, many different fascinating phenomena such as solitons, rogue waves, nonlinear quantum entanglement can be observed.

Some researchers have studied various forms of nonlinear quantum harmonic oscillator (NQHO)  \cite{Kivshar, Carinena2007, Zheng, Ranada, SchulzeHalberg2012, SchulzeHalberg2013}. While some of the research focuses on the effects of nonlinearly behaving spring like potential, the effect of nonlinear wavefunction due to nonlinear electric field is also studied in \cite{Kivshar}. The form of NQHO proposed in \cite{Kivshar} can only be solved numerically unless under some limiting cases \cite{Kivshar}.

In this paper, we propose a numerical framework for studying NQHO. More specifically, we first develop a split-step Fourier scheme for the numerical solution of a more general NQHO than the one proposed in \cite{Kivshar}. We test the accuracy of the proposed scheme using some analytical solutions as benchmark problems which can be derived in the limiting cases. We also test it against a $4^{th}$ order Runge-Kutta solver we developed for the numerical solution of the NQHO. Then, using the split-step Fourier scheme proposed, we analyze the rogue quantum harmonic oscillations, which are unexpected and high amplitude oscillations, in the frame of the NQHO. We show that modulation instability (MI) triggers rogue oscillations to be formed within the frame of the NQHO. We discuss the effects of strengths of trapping well potential, nonlinearity, dissipation, fundamental wave number and perturbation amplitude on rogue oscillation formation probabilities. We discuss and comment on our findings.

\section{\label{sec:level2}A Nonlinear Quantum Harmonic Oscillator Model}

Most common form of the Hamiltonian of the linear QHO (LQHO) can be written as \cite{Schrodinger, Griffiths, Liboff, Pauli, Messiah}
\begin{equation}
\widehat{H}=\frac{\widehat{p}^2 }{2 m}+\frac{1}{2}k x^2=\frac{\widehat{p}^2 }{2 m}+\frac{1}{2}m \omega^2 x^2
\label{eq01}
\end{equation}
where $H$ denotes the Hamiltonian of the system, $m$ is the mass of the particle, $k$ is the bond stiffness which is analogous to spring constant in classical mechanics. $\widehat{p}=-i \hbar \partial / \partial x $ is the momentum operator where $\hbar$ is the reduced Planck's constant. Time-dependent Schr\"{o}dinger equation can easily be derived using this Hamiltonian as
\begin{equation}
i \hbar \frac{\partial \psi}{\partial t}+\frac{\hbar ^2 }{2 m}\frac{\partial^2 \psi}{\partial x^2}-\frac{1}{2}m \omega^2 x^2 \psi=0
\label{eq02}
\end{equation}
which is the most commonly studied form of the linear QHO in the literature \cite{Schrodinger, Griffiths, Liboff, Pauli, Messiah}. In here, $i$ is the imaginary unity, $t$ is the temporal variable, $x$ is the spatial coordinate and $\psi (x,t)$ is the wavefunction, that is the quantum state of the atomic particle at a given $(x,t)$ coordinate. Various forms of NQHO models are considered in \cite{Kivshar, Carinena2007, Zheng, Ranada, SchulzeHalberg2012, SchulzeHalberg2013}. In some of these studies the nonlinearity arises due to nonlinear bonding (spring) constant which results in various forms of the potential function different than the well-studied quadratic form. However, in order to model the effects of nonlinear fields, a NQHO equation is proposed in \cite{Kivshar} which can be written as 
\begin{equation}
i\psi_t +  \frac{\partial^2 \psi}{\partial x^2}- x^2 \psi + \sigma \left|\psi \right|^2 \psi=0
\label{eq03}
\end{equation}
where $\sigma$ is a constant which controls the effect of nonlinearity. It is necessary to note that this equation is a non-dimensional version of the well-known LQHO \cite{Kivshar}. The corresponding non-dimensional parameters also known as Buckhingham's pi terms are given in \cite{Kivshar}. In the linear limit, $\sigma=0$, and this equation reduces to the LQHO which admits solutions in the form of 
\begin{equation}
\psi(x,t)=U(x) e^{-i \omega t}
\label{eq04}
\end{equation}
which only exists for discrete spectrum $\omega_n=1+2n$ where $n=0,1,2,...$ \cite{Kivshar}. And the corresponding amplitude functions become $U_n=(2^n n! \sqrt{\pi} )^{-1/2} e^{-x^2/2}H_n(x)$. Here $H_n(x)$ shows the Hermite polynomials which can be given as
\begin{equation}
H_n(x)=(-1)^n e^{x^2/2} \frac{d^n (e^{-x^2/2})}{dx^n}
\label{eq05}
\end{equation}
giving $H_0=1$, $H_1=2x$, $H_2=4x^2-2$,... etc \cite{Kivshar, Abramowitz, Ryshik}. The solutions of Eq.(\ref{eq03}) are studied using perturbation theory in terms of Hermite-Gaussian eigenmodes and also numerically in \cite{Kivshar}, where a continuous frequency spectrum is considered.

In this paper, we consider a more general version of the NQHO given by Eq.(\ref{eq03}) first proposed in \cite{Kivshar}. In order to take the effects of variable potential due to varying bonding (spring) stiffness, we take a potential well constant $\alpha$ into account, and in order to model the effects of dissipative process or medium, we take a dissipation constant, $\gamma$, into account. Thus, the non-dimensional NQHO equation considered in this paper becomes 
\begin{equation}
i\psi_t +  \frac{\partial^2 \psi}{\partial x^2}-\alpha x^2 \psi + \sigma \left|\psi \right|^2 \psi+ i \gamma \psi =0
\label{eq06}
\end{equation}
where $t$ is the non-dimensional time, $x$ is the non-dimensional space parameter, as before. In the next sections of this paper, we propose a split-step Fourier method for the solution of  Eq.(\ref{eq06}) and we study the rogue quantum harmonic oscillations and their statistics within its frame.

\section{Split-Step Fourier Method for the Numerical Solution of the Nonlinear Quantum Harmonic Oscillator}

In this section, we propose a numerical solution method for the solution of the NQHO given by Eq.(\ref{eq06}). The method we propose is a split-step Fourier method (SSFM). Some applications of various SSFMs for different models can be seen in \cite{BayPRE1, BayTWMS2015, BayTWMS2016}. We first split the NQHO equation to nonlinear and linear parts. A possible splitting gives the nonlinear part of the NQHO which can be written as
\begin{equation}
i\psi_t= -(-\alpha x^2 +  \sigma \left| \psi \right|^2+ i\gamma)\psi
\label{eq07}
\end{equation}
and can be exactly solved to give
\begin{equation}
\tilde{\psi}(x,t_0+\Delta t)=e^{i(-\alpha x^2 + \sigma \left| \psi_0 \right|^2+ i\gamma )\Delta t}\ \psi_0   
\label{eq08}
\end{equation}
In here, $\psi_0=\psi(x,t_0)$ is the initial condition. $\Delta t$, which shows the time step is selected as $\Delta t=5\times 10^{-5}$, which does not cause stability problems, for all of our simulations in this study. The remaining linear part of the NQHO equation can be written as
\begin{equation}
i\psi_t=-\psi_{xx}
\label{eq09}
\end{equation}
It is possible to evaluate this linear part of the NQHO equation in periodic domain spectrally using Fourier series, so that it can be calculated using
 \begin{equation}
\psi(x,t_0+\Delta t)=F^{-1} \left[e^{-i k^2\Delta t}F[\tilde{\psi}(x,t_0+\Delta t) ] \right]
\label{eq10}
\end{equation}
where $k$ is the wavenumber \cite{Bay_CSRM}. Number of spectral components are selected to be $N=1024$ for FFT routines used in all of our simulations in this paper. Therefore pluging Eq.(\ref{eq08}) into Eq.(\ref{eq10}), the complete form of the SSF scheme for NQHO equation can be written as
 \begin{equation}
%\begin{split}
\psi(x,t_0+\Delta t)= F^{-1} \left[e^{-ik^2\Delta t} F[ e^{i(-\alpha x^2+ \sigma \left| \psi_0 \right|^2 +i\gamma  )\Delta t}\ \psi_0 ] \right]
%\end{split}
\label{eq11}
\end{equation}
Starting from the initial condition and using two FFT routines per time step, the numerical solution of the NQHO equation given by Eq.(\ref{eq06}) is obtained for later times by the SSFM for which the scheme is given by  Eq.(\ref{eq11}). 

Initial condition $\psi_0$ can be selected arbitrarily, however in order to study the dynamics of rogue quantum harmonic oscillations we use
 \begin{equation}
\psi_0= e^{i m k_0 x}+ \beta a(x)
\label{eq12}
\end{equation}
as our initial condition. Such an initial condition, in the form of white-noise superimposed to a sinusoid, triggers modulation instability (MI), which causes the formation of rogue quantum harmonic oscillations, similar to those observed in optics, Bose-Einstein condensation, marine environment, finance just to name a few \cite{Kharif, Akhmediev2009b, Akhmediev2009a, Akhmediev2011, BayPRE2, Wang, BayPLA, Birkholz, Soto2014RwSSchaotic, Peregrine, Bay_arxChaotCurNLS, Narhi, Haver, Zhao2013, dqiu, Bay_arxNoisyTunKEE, Bay_arxEarlyDetectCS}. In Eq.(\ref{eq12}), $m$ is an integer for which few different values are considered in our numerical calculations, $k_0=2 \pi /L$ which is the fundamental wavenumber, $\beta$ is the perturbation amplitude and a(x) is a vector of random numbers uniformly distributed in the interval of [-1,1]. We select the domain length as $L=20$ and discuss the effects the parameters $m$ and $\beta$, as well as the $\alpha, \sigma, \gamma$ parameters of the Eq.(\ref{eq06}) on rogue oscillation formation probabilities in the next section.

\section{\label{sec:level3}Results and Discussion}
 
First, we implement the SSFM for the solution of the NQHO using the scheme and parameters as formulated in the preceding section and tested its stability and accuracy against some analytical solutions of the NQHO which can be obtained for limiting cases given in \cite{Kivshar}. Then, we focus on the full form of the NQHO given by Eq.(\ref{eq06})
and compare the accuracy and stability of the scheme against a $4^{th}$ order Runge-Kutta solver we have developed for the NQHO, similar to the ones given in \cite{Canuto, Demiray2015, Karjadi2010, Karjadi2012, BayMS, trefethen, BayScienrep}. 

\begin{figure}[htb!]
\begin{center}
   \includegraphics[width=3.4in]{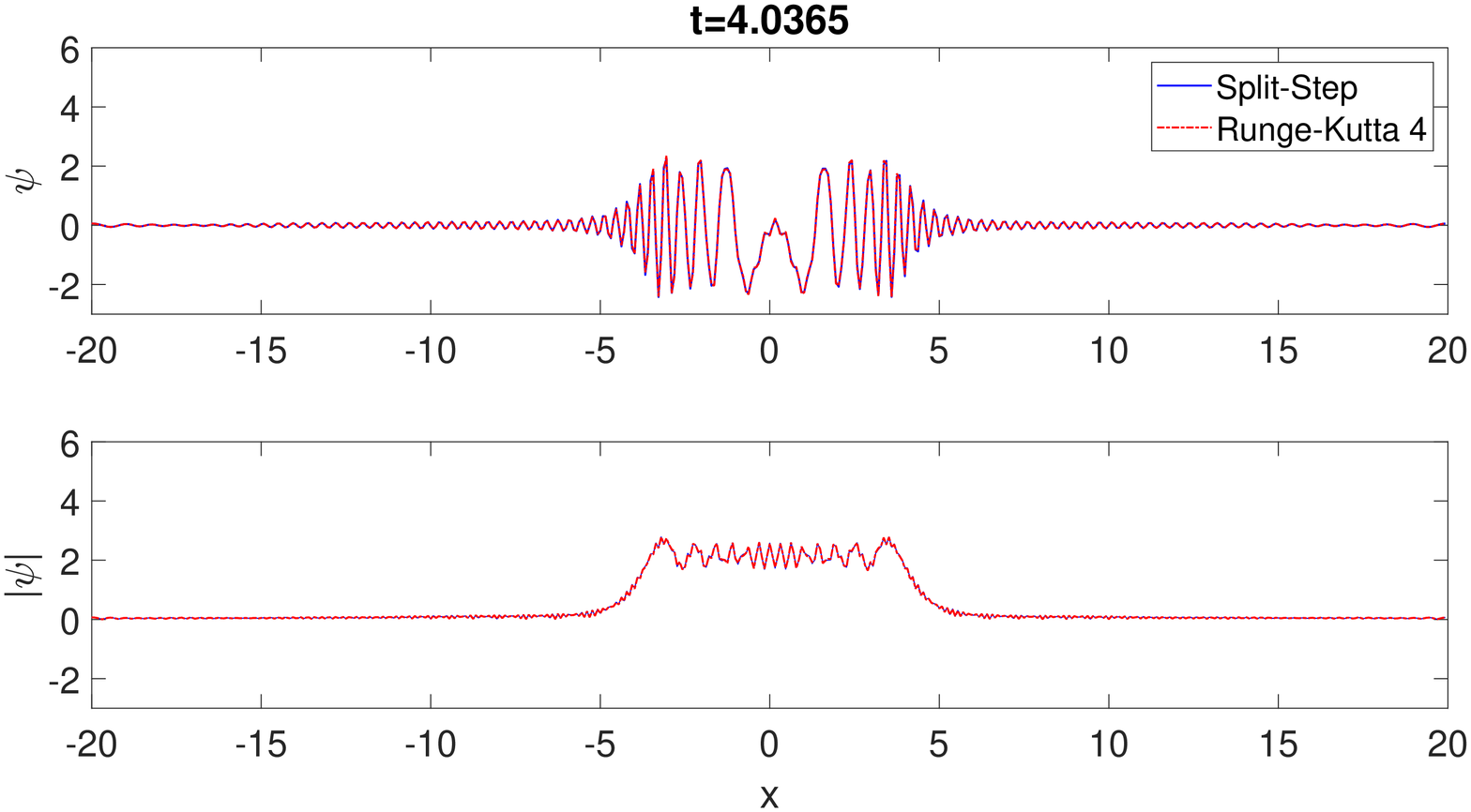}
  \end{center}
\caption{\small Comparison of numerical solutions of the NQHO, SSFM vs. $4^{th}$ order Runge-Kutta scheme. }
  \label{fig1}
\end{figure}
In Fig.~\ref{fig1}, we depict the numerical solutions of the NQHO obtained by SSFM and $4^{th}$ order Runge-Kutta scheme for an initial condition in the form of a sinusoidal without any noise triggering the MI. The parameters of computation are selected as $\alpha=1, \sigma=0, \gamma=0$ for this simulation. Checking Fig.~\ref{fig1} and other test cases for analyzing the accuracy and stability of the proposed method, we conclude that SSFM proposed in this paper can be used for the numerical studies of the NQHO. Our aim is to study the dynamics of rogue oscillations of the NQHO, thus we turn our attention from numerical aspects to the characteristics of rogue oscillations, which can be described as the oscillations having a height at least twice as large as the significant oscillation height in the chaotic wavefield.

\begin{figure}[htb!]
\begin{center}
   \includegraphics[width=3.4in]{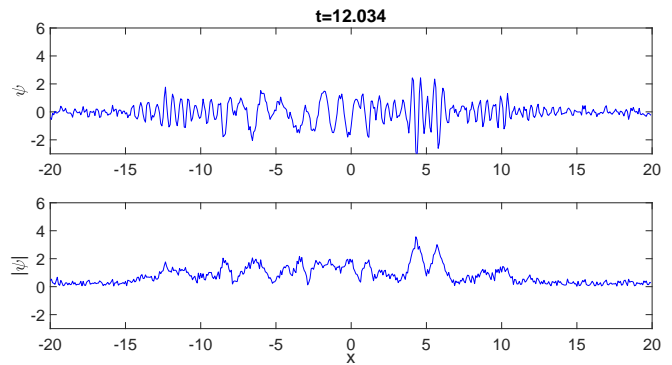}
  \end{center}
\caption{\small A snapshot of the wavefunction exhibiting rogue quantum harmonic oscillations.}
  \label{fig2}
\end{figure}

\begin{figure}[htb!]
\begin{center}
   \includegraphics[width=3.4in]{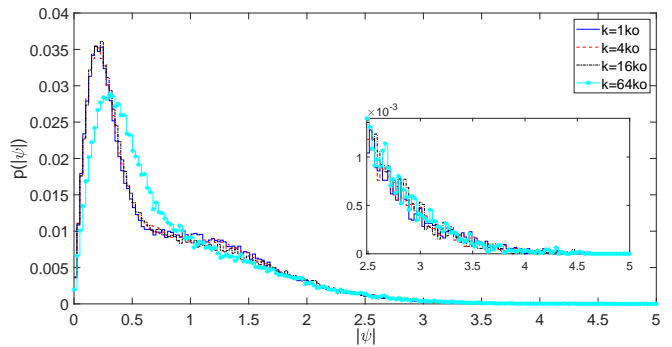}
  \end{center}
\caption{\small Rogue oscillation formation probability for different values of the fundamental wavenumber, $m k_0$.}
  \label{fig3}
\end{figure}

\begin{figure}[htb!]
\begin{center}
   \includegraphics[width=3.4in]{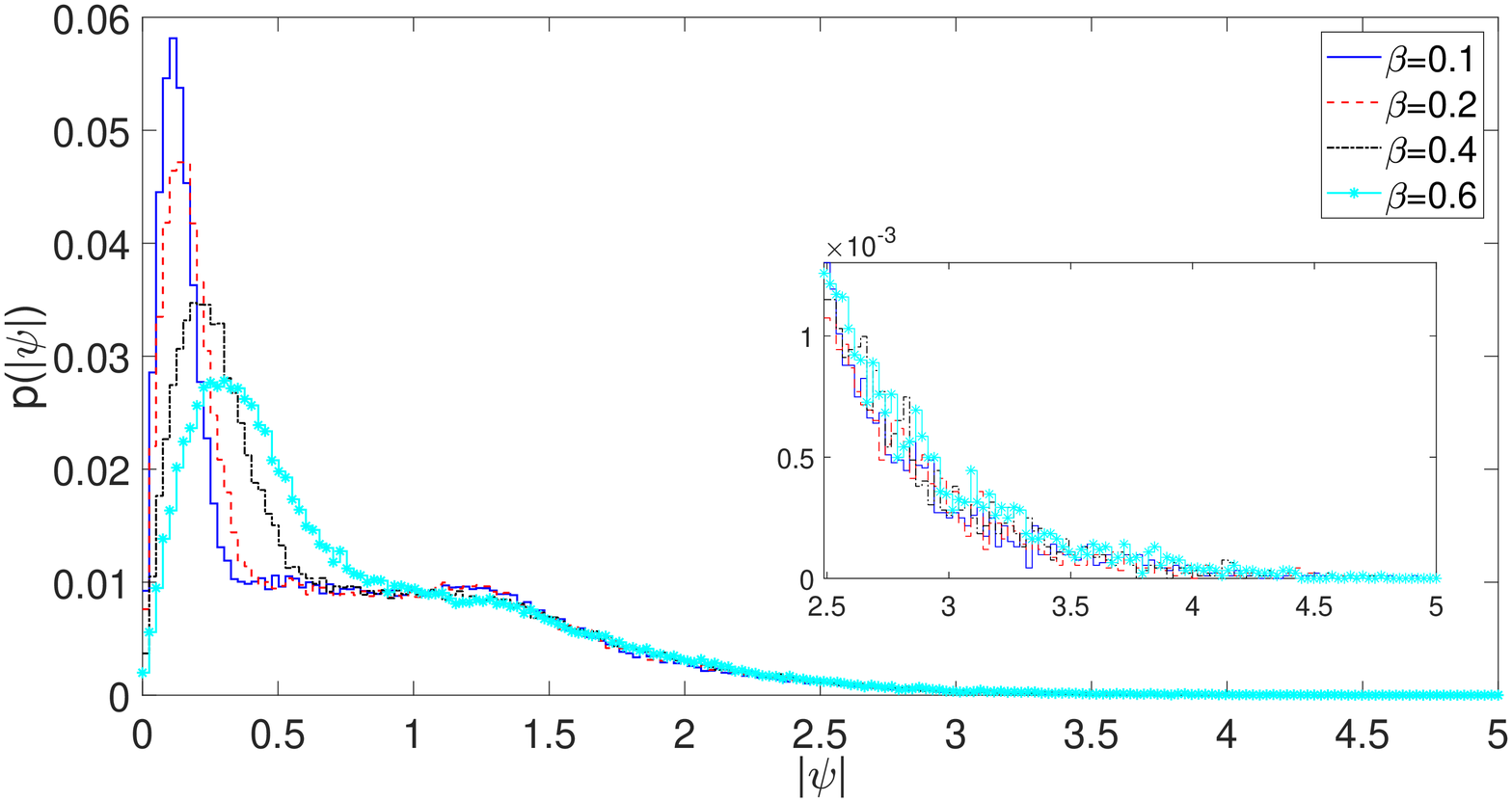}
  \end{center}
\caption{\small Rogue oscillation formation probability for different values of the MI parameter, $\beta$.}
  \label{fig4}
\end{figure}

\begin{figure}[htb!]
\begin{center}
   \includegraphics[width=3.5in]{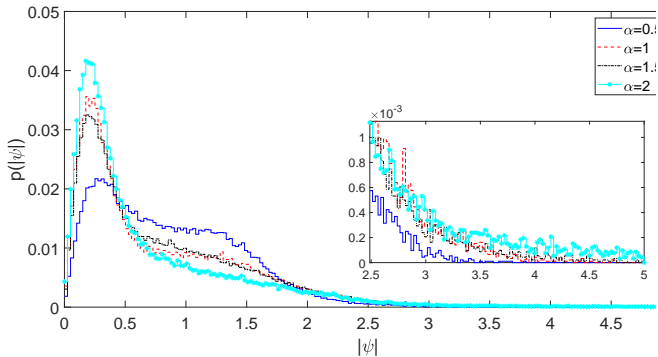}
  \end{center}
\caption{\small Rogue oscillation formation probability for different values of the NQHO trapping well potential strength, $\alpha$.}
  \label{fig5}
\end{figure}

\begin{figure}[htb!]
\begin{center}
   \includegraphics[width=3.4in]{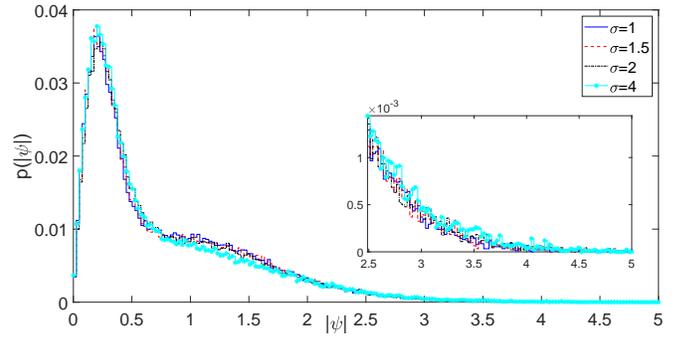}
  \end{center}
\caption{\small Rogue oscillation formation probability for different values of the NQHO nonlinearity parameter, $\sigma$.}
  \label{fig6}
\end{figure}

\begin{figure}[htb!]
\begin{center}
   \includegraphics[width=3.4in]{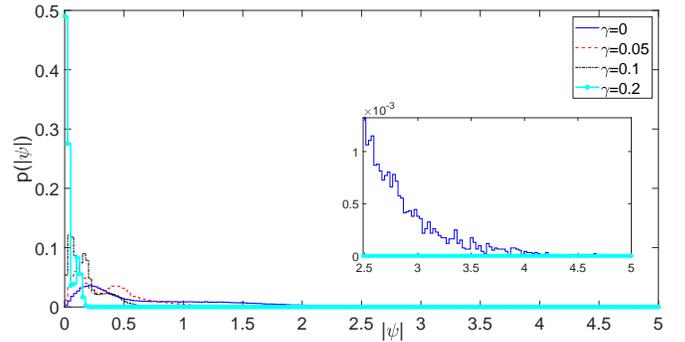}
  \end{center}
\caption{\small Rogue oscillation formation probability for different values of the NQHO dissipation parameter, $\gamma$.}
  \label{fig7}
\end{figure}

In Fig.~\ref{fig2}, a typical wavefield with rogue quantum harmonic oscillations is presented. For this simulation the parameters are selected as $\alpha=1, \sigma=1, \gamma=0, m=16, \beta=0.4$. Due to noisy initial condition and nonlinearity, MI triggers generation of rogue quantum harmonic oscillations within the frame of the NQHO as depicted in Fig.~\ref{fig2}. Although the analytical forms of such rogue oscillations are unknown, their dynamics and statistics are quite similar to the rogue waves discussed in \cite{Akhmediev2009a, BayPRE2}. In order to illustrate this picture we depict Figs.~\ref{fig3}-\ref{fig7}. In order to capture the steady state statistics of the chaotic wavefields, a non-dimensional adjustment time of $t=10$ is given to the SSFM scheme, and statistics are recorded after this adjustment time. Each of statistics depicted in Figs.~\ref{fig3}-\ref{fig7} includes almost $10^5$ wave components to achieve statistically meaningful results.

 Checking Fig.~\ref{fig3}, it can be observed that increase in the fundamental wavenumber of the initial condition, $m k_0$, leads to a higher probability of wavefunctions to be in the range of $\left|\psi\right| \approx [0.4-1.0]$ and a lower probability of them to be in the ranges of  $\left|\psi\right| \approx [0.0-0.4]$ and  $\left|\psi\right| \approx [1.0-2.0]$. The probability of the rogue wavefunctions in the range of $\left|\psi\right|\approx [2.0-5.0]$ can be observed to be almost unaltered due to changes in $k_0$.

Next, we investigate the effect of the MI parameter $\beta$ on the statistics of rogue oscillations by selecting the numerical computation parameters as $\alpha=1, \sigma=1, \gamma=0, m=16$. Checking Fig.~\ref{fig4}, it is possible to state that increase in the MI parameter $\beta$, results in higher rogue oscillation formation probability except for the interval $\left|\psi\right|\approx [1.1-1.5]$. This is due to the fact that a larger $\beta$ feeds the NQHO with more energy and causes more energy transfer from wavenumbers having moderate energy to sidebands.

The effect of the strength of the trapping well potential, that is changing $\alpha$, on rogue oscillation formation probability is depicted in Fig.~\ref{fig5} using $\sigma=1, \gamma=0, m=16, \beta=0.4$. The striking feature of the figure is that, increasing the strength of the trapping well potential leads to increase in low, $\left|\psi\right|\approx [0.0-0.5]$, and rogue, $\left|\psi\right|\approx [2.0-5.0]$, oscillation formation probability. One possible physical reasoning for this fact is that, higher bonding (spring) stiffness in the form of stronger trapping potential imposes a confinement in the oscillations, thus enables to more energy transfer from wavenumber components with low energy to the ones having more energy. Thus, rogue oscillation formation probability increases as $\alpha$ increases. 

Statistics of rogue oscillations for different values of the nonlinearity paramater, $\sigma$, are depicted in Fig.~\ref{fig6} for $\alpha=1,\gamma=0, m=16, \beta=0.4$. Checking the figure, it is possible to conclude that increase in $\sigma$ leads to higher probability for wavefunction to be in the intervals of  $\left|\psi\right|\approx [0.0-0.7]$ and $\left|\psi\right|\approx [1.7-5.0]$, however the probability of wavefunction to be in the interval of $\left|\psi\right|\approx [0.7-1.7]$ decreases as $\sigma$ increases, due to stronger interactions leading to more energy leakage to sideband wavenumbers. A similar result can also be seen in \cite{BayPRE2} for the Kundu-Eckhaus equation which has quintic nonlinearity effect.

Lastly, we investigate the effects of dissipation on the rogue oscillation statistics for which we depict the results in Fig.~\ref{fig7}. Similarly, the numerical values of parameters are selected as $\alpha=1, \sigma=0, m=16, \beta=0.4$ to allow inter comparisons between figures. As expected, increase in the dissipation coefficient, $\gamma$, significantly decreases the rogue and high oscillation formation probabilities. Even a value of $\gamma=0.1$ is strong enough to prevent formation of rogue oscillations in the frame of the NQHO. 

Statistics presented above for various scenarios and the dynamics of the oscillations studied in the frame of the NQHO suggests that the forms of its rogue oscillations may be similar to Peregrine-Akhmediev solitons discussed in \cite{Akhmediev2009a, Peregrine}. The numerical results in our paper and the SSFM we have proposed can be used to investigate nonlinear quantum oscillations and for analyzing resonance and bonding strength of molecules under the effect of nonlinear electric and magnetic fields, dissipation and trapping well potentials having variable strengths. Additionally, our paper can have possible applications in the macroscopic level, such as Bose-Einstein condensation.

\section{\label{sec:level1}Conclusion and Future Work}

In this paper we developed a split-step Fourier scheme for the numerical solution of the nonlinear quantum harmonic oscillator. Although the numerical method proposed can be used to analyze various forms of nonlinear quantum harmonic oscillations, we focused on rogue quantum harmonic oscillations which can be described as high and unexpected oscillations (standing waves). We showed that an initial condition in the form of white noise imposed on monochromatic wave triggers modulation instability, thus leads to chaotic wavefields exhibiting rogue oscillations in the frame of the nonlinear quantum harmonic oscillator. We also discussed the effects of various parameters such as the strength of trapping well potential, nonlinearity, dissipation, fundamental wave number and perturbation amplitude on statistics of rogue oscillations. Our results can be used to investigate nonlinear quantum harmonics oscillations under the effects of varying molecular bond stiffness, higher order nonlinearity and dissipation. Various similar phenomena at the macroscopic level, such as Bose-Einstein condensation, can also be studied analogously using the framework proposed in this paper.


\begin{thebibliography}{00}

\bibitem{Schrodinger}
E. Schrödinger, Annalen d. Physik (4), 79, 489 (1926).

\bibitem{Griffiths}
D. J. Griffiths, \textit{Introduction to Quantum Mechanics} (Prentice Hall, Harlow, 2004).

\bibitem{Liboff}
R. L. Liboff, \textit{Introductory Quantum Mechanics} (Addison–Wesley, New York, 2002).

\bibitem{Pauli}
W. Pauli, \textit{Wave Mechanics: Volume 5 of Pauli Lectures on Physics} (Dover, New York, 2000).

\bibitem{Messiah}
A. Messiah, \textit{Quantum Mechanics}, (North-Holland, Amsterdam, 1967).

\bibitem{Nattermann}
P. Nattermann, Sym. Nonl. Math. Phys.,  {\bf{2}}, 270 (1997).

\bibitem{Reinisch}
G. Reinisch, Phys. A: Stat. Mech. Appl.,  {\bf{206}}, 229 (2001).

\bibitem{Wang_Nl_entang}
G. Wang, L. Huang, Y. C. Lai, and C. Grebogi, Phys. Rev. Lett.,  {\bf{112}}, 110406 (2004).

\bibitem{Bay_Zeno}
C. Bay\i nd\i r and F. Ozaydin. Opt. Commun., {\bf{413}}, 141 (2018).

\bibitem{Bay_TWMS2017}
C. Bay\i nd\i r. TWMS Journal of Applied and Engineering Mathematics,  {\bf{7}}, 236 (2017). 

\bibitem{Chia_Vedral}
A. Chia, M. Hajdušek, R. Fazio, L. C. Kwek and V. Vedral, arXiv Preprint,  arXiv:1711.07376, 2017.

\bibitem{Kivshar}
Y. S. Kivshar, T. J. Alexander and S. K. Turitsyn, Phys. Lett. A,  {\bf{278}}, 225 (2001).

\bibitem{Carinena2007}
J. F. Cariñena, M. F. Rañada and M. Santander, Ann. Phys., {\bf{322}}, 434 (2007). 

\bibitem{Zheng}
L. Zheng, T. Wang, X. Zhang and L. Ma, Appl. Math. Lett.,  {\bf{26}}, 463 (2013).

\bibitem{Ranada}
M. F. Ranada, J. Math. Phys.,  {\bf{55}}, 082108 (2014).

\bibitem{SchulzeHalberg2012}
A. Schulze-Halberg and J. R. Morris, J. Phys. A,  {\bf{45}}, 305301 (2012).

\bibitem{SchulzeHalberg2013}
A. Schulze-Halberg and J. R. Morris, J. Math. Phys.,  {\bf{54}}, 112107 (2013).

\bibitem{Abramowitz}
M. Abramowitz and I. Stegun, \textit{Handbook of Mathematical Functions with Formulas, Graphs, and Mathematical Tables}, (Dover Publications, New York, 1964). 

\bibitem{Ryshik}
I. M. Ryshik and I. S. Gradstein, \textit{Tables of Integrals, Series, and Products}, (Academic Press, New York, 1965).

\bibitem{BayPRE1}
C. Bay\i nd\i r. Phys. Rev. E,  {\bf{93}}, 032201 (2016).

\bibitem{BayTWMS2015}
C. Bay\i nd\i r. TWMS Journal of Applied and Engineering Mathematics,  {\bf{5-2}}, 298 (2015).

\bibitem{BayTWMS2016}
C. Bay\i nd\i r. TWMS: Journal of Applied and Engineering Mathematics,  {\bf{6-1}}, 135 (2016).

\bibitem{Bay_CSRM}
C. Bay\i nd\i r. TWMS: Journal of Applied and Engineering Mathematics,  {\bf{8-2}}, 425 (2018). 

\bibitem{Kharif}
C. Kharif and E. Pelinovsky. European Journal of Mechanics, B: Fluids.  {\bf{6}}, 603 (2003).

	\bibitem{Akhmediev2009b}
N. Akhmediev, A. Ankiewicz and J. M. Soto-Crespo, Phys. Rev. E,  {\bf{80}}, 026601  (2009).

\bibitem{Akhmediev2009a}
N. Akhmediev, J. M. Soto-Crespo and A. Ankiewicz, Phys. Lett. A,  {\bf{373}}, 2137 (2009).

\bibitem{Akhmediev2011}
N. Akhmediev, J. M. Soto-Crespo, A. Ankiewicz and N. Devine, Phys. Lett. A,  {\bf{375}}, 2999  (2011).

\bibitem{BayPRE2}
C. Bay\i nd\i r. Phys. Rev. E,  {\bf{93}}, 062215 (2016).

\bibitem{Wang}
X. Wang, B. Yang, Y. Chen and Y. Yang. Phys. Scr.,  {\bf{89}}, 095210 (2014).

\bibitem{BayPLA}
C. Bay\i nd\i r. Physics Letters A,  {\bf{380}}, 156 (2016).

\bibitem{Birkholz}  S. Birkholz, C. Bree, A. Demircan and G. Steinmeyer. Phys. Rev. Lett., {\bf{114}}, 213901, 2015.

\bibitem{Soto2014RwSSchaotic}
J. M. Soto-Crespo, N. Devine, N.P. Hoffmann and N. Akhmediev. Phys. Rev. E,  {\bf{90}}, 032902 (2014).

\bibitem{Peregrine}
D. H. Peregrine, J. Austral. Math. Soc. B.,  {\bf{25}}, 16 (1983).

\bibitem{Bay_arxChaotCurNLS}
C. Bay\i nd\i r. arXiv Preprint,  arXiv:1512.03584 (2016).

\bibitem{Narhi} M. Narhi, L. Salmela, J. Toivonen, C. Billet, J. M. Dudley and Goëry Genty. Nature Comm., {\bf{9}}, 4923, 2018.

\bibitem{Haver} S. Haver, Proc. Rogue Waves, Ifremer, 129, 2000.
 
\bibitem{Zhao2013}     
L. C. Zhao, C. Liu and Z. Y. Yang. Communications in Nonlinear Science and Numerical Simulation,  {\bf{20-1}}, 9 (2015).

\bibitem{dqiu}
D. Qiu,  J. He, Y. Zhang and K. Porsezian. Proceedings of the Royal Society A,  {\bf{471}}, 20150236 (2015).

\bibitem{Bay_arxNoisyTunKEE}
C. Bay\i nd\i r. 12th International Congress on Advances in Civil Engineering, Istanbul, Turkey (2016).  (arXiv Preprint,  arXiv:1602.05339).

\bibitem{Bay_arxEarlyDetectCS}
C. Bay\i nd\i r. TWMS: Journal of Applied and Engineering Mathematics,  {\bf{9-1}}, (2019). (to be published) (arXiv Preprint,  arXiv:1602.00816 (2016)).

\bibitem{Canuto}
C. Canuto. \textit{Spectral Methods: Fundamentals in Single Domains} (Springer-Verlag, 2006).

\bibitem{Demiray2015}
H. Demiray and C. Bay\i nd\i r. Phys. of Plasm.,  {\bf{22}}, 092105 (2015).

\bibitem{Karjadi2010}
E. A. Karjadi, M. Badiey and J. T. Kirby. The Journal of the Acoustical Society of America,  {\bf{127}}, 1787 (2010).

\bibitem{Karjadi2012}
E. A. Karjadi, M. Badiey, J. T. Kirby and C. Bay\i nd\i r. IEEE Journal of Oceanic Engineering,  {\bf{37-1}}, 112 (2012).

\bibitem{BayMS}
C. Bay\i nd\i r. MS Thesis, University of Delaware, (2009).

\bibitem{trefethen}
L. N. Trefethen. \textit{Spectral Methods in {MATLAB}}, (SIAM, 2000).

\bibitem{BayScienrep}
C. Bay\i nd\i r. Sci. Rep.,  {\bf{6}}, 22100 (2016).


\end{thebibliography}
\end{document}